\documentclass[twocolumn,apj]{aastex631}
\usepackage{amsthm,amsmath,amssymb}
\usepackage{url}
\usepackage{mathrsfs}
\usepackage{hyperref}
\usepackage{mathtools}

\def\reflionx{\textsc{reflionx}}
\def\xillver{\textsc{xillver}}
\def\relxill{\textsc{relxill}}
\def\xstar{\textsc{xstar}}
\newcommand {\bc}{\begin {center}}
\newcommand {\ec}{\end {center}}
\newcommand {\be}{\begin {equation}}
\newcommand {\ee}{\end {equation}}
\newcommand {\beq}{\begin {eqnarray}}
\newcommand {\eeq}{\end {eqnarray}}

\def\xmm{\textit{XMM-Newton}}
\def\nustar{\textit{NuSTAR}}

\shorttitle{High Density Reflection Model}
\shortauthors{Ding et al.}
\graphicspath{{./}{figures/}}

\begin{document}

\correspondingauthor{Yuanze Ding}
\email{yding@caltech.edu}

\title{Next Generation Accretion Disk Reflection Model: High-Density Plasma Effects}

\author{Yuanze Ding}
\affiliation{Division of Physics, Mathematics and Astronomy, California Institute of Technology, Pasadena 91125, United States}
	
\author{Javier A. García}
\affiliation{X-ray Astrophysics Laboratory, NASA Goddard Space Flight Center, Greenbelt 20771, USA}
\affiliation{Division of Physics, Mathematics and Astronomy, California Institute of Technology, Pasadena 91125, United States}

\author{Timothy R. Kallman}
\affiliation{X-ray Astrophysics Laboratory, NASA Goddard Space Flight Center, Greenbelt 20771, USA}

\author{Claudio Mendoza}
\affiliation{Physics Center, Venezuelan Institute for Scientific Research (IVIC), Caracas 1020, Venezuela}

\author{Manuel Bautista}
\affiliation{Department of Physics, Western Michigan University, Kalamazoo, MI 49008, USA}

\author{Fiona A. Harrison}
\affiliation{Division of Physics, Mathematics and Astronomy, California Institute of Technology, Pasadena 91125, United States}

\author{John A. Tomsick}
\affiliation{Space Sciences Laboratory, 7 Gauss Way, University of California, Berkeley 94720, United States}

\author{Jameson Dong}
\affiliation{Division of Physics, Mathematics and Astronomy, California Institute of Technology, Pasadena 91125, United States}

\begin{abstract}
Luminous accretion disks around black holes are expected to have densities of $\sim 10^{15-22}$\,cm$^{-3}$, which are high enough such that plasma physics effects become important. Many of these effects have been traditionally neglected in the calculation of atomic parameters, and therefore from photoionization models, and ultimately also from X-ray reflection models. In this paper, we describe updates to the atomic rates used by the \xstar\ code, which is in turn part of the \xillver\ disk reflection model. We discuss the effect of adding necessary high density corrections into the \xillver\ code. Specifically, we find that the change of recombination rates play an important role, dominating the differences between model versions. With synthetic spectra, we show that even in a highly ionized state, high density slabs can produce strong iron ($\sim$6.5-9\,keV) and oxygen ($\sim0.6-0.8$\,keV) resonance features. The significant iron emission could address the problem of the supersolar iron abundances found in some sources.
\end{abstract}


\section{Introduction}
\label{sec:Intro}

The reflection spectrum from accreting black holes (BHs) is thought to be produced in a relatively cold accretion disk, being illuminated by copious hard X-rays from a hot corona in the vicinity of the BH \citep[][]{thorne1975,lightman1988,haardt1991,haardt1993}. The spectrum shows similar features both in black hole X-ray binary (BHXRB) and active galactic nuclei (AGN): iron K lines at $\sim6.4-6.9$\,keV and absorption K edge just beyond the iron K peak, together with a broad Compton hump at $\sim$25\,keV \citep{fabian1989,Matt1991A&A,laor1991}. The former is a prominent emission component in the group of X-ray fluorescent lines, and the latter is a Compton scattered continuum formed into the shape of a hump with the range determined by atomic absorption on the red side and electron scattering on the blue side. These are the most obvious and relevant features of reflection, though other astrophysically abundant atoms (e.g., nitrogen, oxygen, sulfur, silicon and calcium) present in the accretion disk can also produce spectral features in emission or absorption. The blend of these lines are considered to be a possible origin of the soft excess in AGN \citep{garcia2016effects, garcia2019_mrk509}. These features can be further distorted by the strong gravitational field around BH, thus providing important information on the BH itself.


In recent decades, there have been numerous efforts to develop accretion disk reflection models \cite[see][for a review of theory and modeling method]{bambi2021towards}. Traditional accretion disk reflection models assume a fixed density through out the solving region, typically $n_{e}=10^{15}\,$cm$^{-3}$ \citep{ross2005,garcia2010,garcia2013}. This approximation greatly simplifies the calculations required. However, for low mass AGN ($10^{5}-10^{7}\,M_{\odot}$) and BHXRBs, it has been theoretically predicted that the inner accretion disk can achieve higher densities than the values typically assumed \citep{shakura1973,Svensson1994ApJ}. \cite{garcia2016effects} summarized some of the effects of the gas density on the X-ray reflection spectra, using the state-of-the-art reflection model \relxill. However, the only relevant high-density effect in this model is the change of free-free emissivity and opacity, while other plasma effects {/bf associated with high densities} were ignored.

Observationally, the zeroth-order high-density effects like those described in the reflection models by \cite{garcia2016effects} have been suggested as an alternative explanation for the extreme super-solar iron abundances measured through X-ray reflection spectroscopy in several BHXRBs and AGNs \cite[e.g.,][]{garcia2018, Tomsick2018ApJ,Jiang2018high,Jiang2022mnras}. In the case Cyg~X-1, it was found that fits with the high-density model recovered values for the iron abundance ($A_\mathrm{Fe}$) that are more realistic \citep[][i.e., equal to Solar]{Tomsick2018ApJ}. A later systematic analysis being conducted by \cite{jiang2019high} revealed that $\sim$65\% of Seyfert 1 AGN may have disk density significantly higher than $n_{e}=10^{15}\,$cm$^{-3}$, with generally lower iron abundances obtained using high density disk reflection models. Both \citet{mallick2022high} and \citet{jiang2019high} found that the black-body-like soft X-ray excess can be well described by the relativistic reflection from such an ionized, higher density disk.

The iron abundance is the only free parameter available to account for uncertainty in the line reprocessing efficiency in state-of-the-art reflection models. Consequently, the super-solar abundances, and apparent correlation between spin and iron abundance \citep{Reynolds2012, Steiner2012MNRAS}, suggest that current models underestimate the strength of the iron line. These are possibly the most straightforward indicators showing that there are systemic uncertainties in the models. An apparent issue in present high-density reflection models is the limited treatment of atomic physics in the high-density regime. In particular, currently available \xillver\ models employ the atomic routines from \xstar\ \cite[2.2.1bn,][]{kallman2001photoionization}, which is only appropriate for densities below $n_{e}\leq10^{18}$cm$^{-3}$.

In this paper, we present accretion disk reflection models including corrections and approximations that are appropriate for electron density ($n_{e}$) up to $\sim10^{22}\,\mathrm{cm}^{-3}$. These include a comprehensive collection of atomic data for such densities which have been calculated since the release of the first publicly available \xillver\ version \cite[see][]{mendoza2021,kallman2021photoionization}. We show that even for density of $n_{e}\simeq10^{15}$cm$^{-3}$, the relatively small changes in atomic parameters can be amplified through the radiative transfer solution of an optically-thick atmosphere, resulting in significant flux changes around iron K edge.


This paper is organized as follows: Section~\ref{sec:method} describes the new mechanisms we include compared to previous models and the impacts on atomic rates; Section~\ref{sec:ion-solution} illustrates the new temperature and ionization solutions and the impact on ion fractions; Section~\ref{sec:decomp} shows tests that are helpful for understanding the differences between code versions; Section~\ref{sec:modelfit} shows how differently the new reflection models describe accretion disk X-ray spectra, via the analysis of simulated observations; Section~\ref{sec:refl-comp} summarizes the difference between \xillver\ and \reflionx\ by comparing the generated spectra; Finally, Section~\ref{sec:conclusion} presents the main conclusions of this paper.

\begin{deluxetable*}{cclcll}[t]
\tablenum{1}
\caption{Summary of the various \xillver, atomic codes, and ATDB versions used in this paper.}
\setcounter{table}{1}
\tablehead{
\colhead{Model \#}&
\colhead{\xillver} &
\colhead{ATDB Creation Date} &
\colhead{ID} &
\colhead{\xstar} & 
\colhead{Notes}
}
\startdata
1 & 3.4b & 2012-08-03T01:58:54 & ATDB12 & \xstar~2.2.1bn & Current release (old codes, old ATDB)\\
2 & 3.5  & 2023-07-17T19:39:30 & ATDB23 & \xstar~2.59    & Full version, new release (new codes, new ATDB)\\
3 & 3.5  & 2023-07-17T19:39:30 & ATDB23 & \xstar~2.59    & Same as Model~2, but including REX \\
4 & 3.5  & 2012-08-03T01:58:54 & ATDB12 & \xstar~2.59    & New codes, but same ATDB as in Model~1 \\
5 & 3.5  & 2021-08-03T18:36:21 & ATDB21 & \xstar~2.59    & New codes, updated ATDB, but incorrect Ca cross sections\\
\enddata
\tablecomments{
Col. (1): Model number. Model~1 and~2 correspond to standard \xillver~3.4b and 3.5. Model~1 is the same as the public version of \xillver\ below $\sim\,$40\,keV (see details in Section~\ref{sec:algorithm of Xillver}).
Col. (2): \xillver\ model version. REX: radiative excitation.
Col. (3): Creation time for the ATDB, typically used as an identifier independent of the model version. 
Col. (4): ATDB ID for this paper.
Col. (5): Version of the \xstar\ routines used. 
\xillver~3.5 uses modified ATDBs in the high-density regime ($n>10^{18} \rm cm^{-3}$), because the DR and CL effects are implemented through the atomic data itself.
Col. (6): Additional notes.}
\label{tab:code_versions}
\end{deluxetable*}

\section{Reflection Modeling at High Density}
\label{sec:method}

In this Section we describe the main components required to model X-ray reflection from accretion disks, namely,
the main code that computes the disk reflection (or reprocessing), and solving radiation transport---the \xillver\ code; the 
routines used for the calculation of the ionization state of the gas---the \xstar\ code; and the atomic database (ATDB) that
describes the microphysics of the problem---the \xstar\ ATDB.

\subsection{Disk reflection: the \xillver\ Model} 
\label{sec:algorithm of Xillver}
The \xillver\ code calculates the radiative transfer of the X-rays incident in an optically-thick disk atmosphere, which are then
reprocessed and reflected back to the observer.  
This code has been described extensively in previous publications \citep{garcia2010,garcia2013,garcia2014}. The code solves the radiative transfer equation in a plane-parallel (slab) geometry using the classical Feautrier method \citep{mihalas1978stellar}. The underlying \xstar\ routines solve for atomic level and thermal equilibrium, while maintaining charge neutrality. Note that the input number density ($n$) for \xillver\ is hydrogen nucleus density. Inside \xillver, it is initialized to $n_{e}$ by multiplying a factor of 1.2. The current model utilizes the Comptonization treatment from \citet{garcia2020accurate}\footnote{New tables of reflected spectra including the new Comptonization treatment are currently underway, no yet made publicly available.}. This is an update compared with the classical Gaussian redistribution kernel assumed in \citet{ross2005} and \citet{garcia2013}. The solving region is a thin layer in the disk surface, determined by a Thomson optical depth $10^{-4}<\tau_{T}<10$. The upper boundary condition is placed at $\tau_{T}=10^{-4}$, while for the lower boundary we assume there is no upward flux. This setup ignores the disk intrinsic thermal emission, or assumes that it is negligible in comparison with the coronal power-law incident at the top. This is a reasonable approximation for AGN and hard state BHXRB disks \citep{garcia2013}, but it might cause non-negligible systematic uncertainty in spin measurements when fitting disk dominant systems, e.g., BHXRBs in the soft state.

Another important assumption in our model is that the gas density throughout the solving region is constant. Such assumption significantly simplify the calculation, and have proven to be sufficient to describe the observational data
\citep{Nayakshin2001ApJ,Rozanska2002MNRAS,Ballantyne2004ApJ}. While some of these works have discussed reflection in an atmosphere under hydrostatic equilibrium, its implementation is computationally much more demanding, while it is still unclear whether a hydrostatic solution provides a more realistic realistic prediction for radiation-pressure dominated, thin accretion disks \citep{shakura1973}. Therefore, in this work we do not consider atmospheres with density gradients in the vertical direction, but instead focus on the impact of changing density value over a wide range.

Hereafter, we refer to the version of the \xillver\ code presented in this paper as \xillver~3.5, with the previous version being \xillver~3.4b. \xillver~3.5 is more flexible than its earlier incarnation, and has been tested with different atomic database (ATDB) and parameter setup, as tabulated in Table~\ref{tab:code_versions}.

\subsection{Photoionization: The \xstar\ Model}

The \xillver\ code uses opacities and emissivities which are calculated by \xstar. \xstar\ calculates the ionization balance, temperature, and excited level populations under the assumption of time-steady equilibrium, including all ions and elements with $Z\leq$30. There have been various updates to the \xstar\ calculation since the original development of the \xillver\ model by \cite{garcia2010} and \cite{garcia2013}.   

One useful way to characterize our models is the ionization parameter:
\begin{equation}
    \xi=\frac{4\pi F_{x}}{n_{e}},
    \label{eqn:xi}
\end{equation}
\noindent
where $F_{x}$ is the net ionizing flux integrated in the $0.1-10^3$\,keV range.\footnote{Note that this differs from the energy range used in \xstar, which is $1-10^3$\,Ry.} The parameter $\xi$ indicates the degree of ionization in gas, and basically scales as the ratio between photo-ionization ($\propto F_{x}n_{e}$) and recombination rate ($\propto n_{e}^{2}$). $\xi$ is expected to characterize our constant density layers to a good approximation if photo-ionization and recombination are the dominant processes that control the thermal equilibrium and ionization balance.  This has been proven to be invalid in high-density atmospheres, as other mechanisms like free-free heating/cooling become important \citep{garcia2016effects}.

The models presented in \citet{garcia2013} (\xillver~3.4b) used \xstar\ version 2.2.1bn, in which the treatment of the atomic processes follows \citet{kallman2001photoionization}. These calculations include the effect of high densities up to $\sim10^{18}\,$cm$^{-3}$. However, the code we present here (\xillver~3.5) uses \xstar\ version 2.59.

\subsection{Microphysics: The \xstar\ Atomic Database (ATDB)}

The \xstar\ atomic database is stored and tabulated separately from the \xstar\ code. Typically, new updates to the database are released simultaneously with new releases of \xstar. However, this convention is not always applied, as code and atomic data modifications can also occur independently. Thus, while \xstar\ code follows a version number control, the atomic database is identified by its creation date. In particular, \xillver~3.4b was designed to use ATDB12 or prior (see Table~\ref{tab:code_versions}), while \xillver~3.5 has been made to compatible with both ATDB23 and ATDB12.  The division between the \xstar\ code and its database is intended to maintain backward compatibility so that new versions of the code can run older versions of the database.  The converse does not generally work, since newer versions of the database are considerably larger, and thus require more memory allocation than was used in the older code versions. 

\label{sec:micro-physics}
Here, we briefly review the key ingredients added in \xstar~2.59 and ATDB23 which we use in this paper \cite[see][for a complete description]{kallman2021photoionization,atdb}. These ingredients include  the addition of a variety of mechanisms important in high-density plasmas, extending the model described in \citet{garcia2013} up to densities of $10^{22}\,$cm$^{-3}$, and thus superseding the models presented in \citet{garcia2016effects}. These processes include:

\begin{itemize}
    \item Stimulated processes: an incoming photon of a specific frequency can interact with an excited ion, causing it to drop to a lower energy level. In a photo-ionized plasma, the incident photon flux can be very large if the density is high and the ionization parameter is the conventional value for the inner accretion disk. Such high photon fluxes can lead to large enhancements in the recombination rate via stimulated recombination. For recombination and radiative decay, rates are enhanced by a factor $1+F_{\epsilon}/\frac{2\epsilon^{3}}{h^{3}c^{2}}$, where $F_{\epsilon}$ is monochromatic intensity at energy $\epsilon$; $h$ and $c$ are Planck's constant and the speed of light. \xstar\ calculates stimulated recombination only for radiative recombination.  It simply includes the `Einstein B' coefficient for each recombination onto a spectroscopic level, by adding a term proportional to the local mean intensity inside the Milne integral.

    \item Suppression of Dielectric Recombination (DR): DR occurs when an electron is captured into an auto-ionizing state of the recombined ion. Its rate depends on the gas density, and would be significantly suppressed in a high-density plasma. \cite{Nikolic2013ApJ...768...82N} provide convenient expressions for DR suppression up to $n_{e}=10^{20}\,$cm$^{-3}$. At densities greater than $10^{20}\,$cm$^{-3}$, DR is almost negligible, so we apply the value at $10^{20}\,$cm$^{-3}$ to all higher density cases. The DR suppression would reduce recombination rates in photoionized plasmas, keeping the gas more ionized. This effect is now implemented through the ATDB and the \xstar~2.59 code.

    \item Continuum Lowering (CL): the crowded environment in a high-density plasma can perturb or unbind the atomic states with high principal quantum numbers (\textit{n}, only in this paragraph). This reduces the number of total states when they are summed to calculate the rate coefficients for recombination, or inhibit any other process where high-\textit{n} states are important. In ATDB12 we only considered the high-\textit{n} cut-off for levels higher than our chosen set of spectroscopic levels, i.e. $n\geq$ 4 or 5 typically. In ATDB23 we used detailed atomic structure calculations, including the effects of Debye screening. This results in simple rules for the energy shift of both bound levels and electron continuum levels \citep[see][for details]{kallman2021photoionization}. For densities greater than 10$^{18}$ cm$^{-3}$ the low density \xstar\ level list is truncated for levels above the lowered electron continuum. One other thing being taken into account at the same time is the shift of the energy levels, which manifests itself as a modest redshift of some lines.  Because these effects are dependent on the plasma screening parameter, which is a function of both temperature and density ($\mu = 1/\lambda_D = \sqrt{4\pi n_e/kT}$), we prepare separate database files for different values of $\mu$ (effectively, for different densities), and in each run we read the corresponding database according to the density input. One caveat is that above densities of $10^{18}\,$cm$^{-3}$, these atomic parameters still assume $T=10^{7}\,$K when evaluating the lowering effect with Debye–H\"uckel theory, as it is quite cumbersome to implement the temperature dependency at the same time (this would require extensive modifications of the \xstar\ routines). We defer this for future work. Generally CL lowers the reflected continuum flux, as suggested by its name.

    \item The atomic data for odd-$Z$ elements and trace iron peak elements has been completely recalculated and updated. The updates in ATDB include radiative and collisional rates for the odd- $Z$ elements below $Z=20$, as well as other trace elements above $Z=20$ \citep{mendoza2017k,mendoza2018k,palmeri2012atomic,palmeri2016k}. The database in this work was based on more accurate calculation methods like Hartree--Fock Relativistic \citep{cowan1981alamos}, Auto-structure \citep{badnell1993electric} and multi-configuration Dirac--Fock \citep{grant1980atomic} codes, with the supplementation of $\sim 10^{4}$ new lines and levels from all ions with three or more electrons of F, Na, P, Cl, K, Ti, V, Mn, Cr, Co, Cu and Zn \citep{kallman2021photoionization}. Generally, the current ATDB is about a factor of 2 larger than its previous incarnation.

    \item We have also found and fixed various minor errors: \xillver's thermal equilibrium treatment of the iron unresolved transition array (UTA) was inaccurate, which we solved in \xillver-3.4b by removal of iron heating rate. We also found a long-standing issue in one of the K shell photoionization cross-sections for Ca~{\sc xv/xiv} was much greater than the expected values, due to a transcription error.  This results in unphysical absorption above 10 keV for a certain narrow range of ionization parameter. Additional minor bugs in \xstar\ 2.2.1bn are summarized in the \xstar\ issues page\footnote{\url{http://heasarcdev.gsfc.nasa.gov/lheasoft/issues.html}}.

\end{itemize}

 Although \xstar\ embodies a relatively complete and up-to-date treatment of the atomic processes, \xillver\ has been using the version ATDB12 and the corresponding \xstar\ version 2.2.1bn since its release in 2013. Calculating reflection at a density of $n>10^{18}\,$cm$^{-3}$ therefore involves serious approximations because many quantities in ATDB12 have been tabulated up to $n=10^{18}\,$cm$^{-3}$, including recombination into high atomic levels. ATDB21 has all these aforementioned corrections included but leaves a calcium cross-section issue unfixed (for illustrative purpose, see Section~\ref{sec:decomp} for details). We corrected all known issues in ATDB23, which is the currently recommended one.

 In the updated \xillver\ code (Model~2) we remove the possibility for including the bound-bound continuum radiative excitation (REX) calculation, as it cannot be treated accurately at this stage (note that in \xstar\ this is controlled by the covering fraction parameter).  Accurate treatment of this process requires a very fine spatial grid since the path length for photons in strong lines is much shorter than the other relevant length scales. The removal of REX has limited impact on the result even for some of the most extreme cases, though it could play a significant role for O \textsc{VIII} 653\,eV Ly$\alpha$ resonant transition when the density is very high (e.g., $n=10^{20}$cm$^{-3}$; see Section~\ref{sec:decomp}). In this work, \xillver~3.5, the updated \xillver\ version, always uses \xstar\ 2.59 (see Table~\ref{tab:code_versions}).

\section{Discussion}
\label{sec:discussion}

\subsection{Ionization and Temperature Profile}
\label{sec:ion-solution}

In order to test the new version of the \xillver\ code, we have produced reflection calculations for different configurations of codes and atomic database versions, as shown in Table~\ref{tab:code_versions}, varying the density and the ionization parameter, while keeping the other model parameters fixed to the same values. We show other relevant model parameters in Table~\ref{tab:model_parameters}.

\begin{deluxetable*}{ccccc}[t]
\tablenum{2}
\caption{Default parameters in model calculations.}
\setcounter{table}{1}
\tablehead{
\colhead{\begin{tabular}[c]{@{}l@{}}Incident \\ Angle \\ ($^{\circ}$)\end{tabular}}&
\colhead{\begin{tabular}[c]{@{}l@{}}ionizing \\ continuum\end{tabular}} &
\colhead{\begin{tabular}[c]{@{}l@{}}Coronal \\ electron \\ temperature \\ (keV)\end{tabular}} &
\colhead{\begin{tabular}[c]{@{}l@{}}Seed \\ blackbody \\ temperature \\ (keV)\end{tabular}} &
\colhead{\begin{tabular}[c]{@{}l@{}}Powerlaw \\ photon \\ index \\ ($\Gamma$)\end{tabular}} 
}
\startdata
55 & \textsc{nthcomp} & 100 & 0.1 & 2.0 \\
\enddata
\tablecomments{
Col. (1): Incident angle of the ionizing continuum flux.
Col. (2): The model assumed for ionizing continuum.
Col. (3): Coronal electron temperature assumed in the ionizing continuum model. 
Col. (4): Seeding blackbody temperature assumed in the ionizing continuum model.
Col. (5): Powerlaw photon index that characterized the asymptotic slope of the ionizing continuum model. Tabulated are the default parameters when calculating all reflection models discussed in this paper unless specified separately.}
\label{tab:model_parameters}
\end{deluxetable*}

The ionizing flux is described by the thermal Comptonization model \textsc{nthcomp} \citep{zdziarski1996broad,1999MNRASZzycki}, where we assume a 0.1\,keV blackbody\footnote{Note that this value is 10 times larger than the one used in public \xillver\ tables, in order to match that used in the \reflionx\ model.} being inverse Comptonized by a 100$\,$keV hot corona. The former determines the location of the low-energy roll over while the latter determines the high-energy one. The part between high and low energy roll over is parameterized by an asymptotic power-law photon index $\Gamma=2$. 

\begin{figure*}
    \centering
    \includegraphics[width=0.8\textwidth]{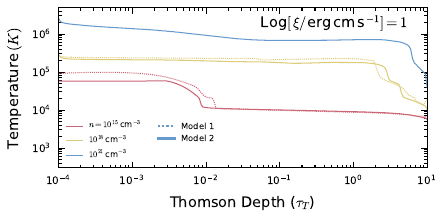}
    \includegraphics[width=0.8\textwidth]{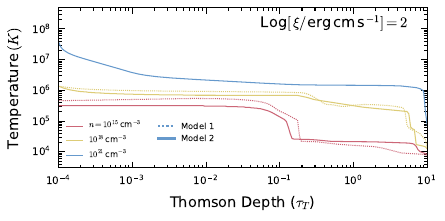}
    \includegraphics[width=0.8\textwidth]{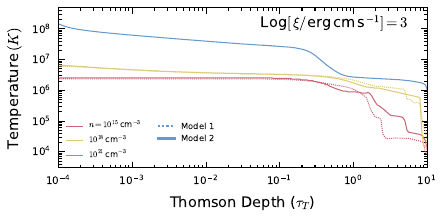}
    \caption{Temperature profiles of the illuminated slab as a function of the Thomson optical depth ($\tau_{T}$). Model~2 (using \xillver~3.5) in the high-density regime generally produce a lower temperature solution. Model~1 (using \xillver~3.4b) cannot iterate correctly when $n>10^{20}\,$cm$^{-3}$, which is due to the erroneous calcium cross-sections.}
    \label{fig:temperature profiles}
\end{figure*}

Figure~\ref{fig:temperature profiles} shows the temperature profile in the vertical direction obtained with three different ionization parameters and densities. The location within the slab is indicated in terms of the Thomson optical depth ($\tau_{T}$). At densities greater than 10$^{18}$\,cm$^{-3}$, Model~1 cannot self-consistently produce a physical temperature solution due to the limitations of the CL calculation and the erroneous calcium cross-section. The corresponding spectra are also incorrect, having no narrow features. We thus ignore the spectrum of $n=10^{21}\,$cm$^{-3}$ for Model~1 in all figures.

With increasing ionization ($\xi$), the profiles gradually coincide near the surface of the slab, where elements are highly ionized. The overall temperature of a given model increases with density. For each model, the temperature profile is similar to those reported in previous calculations at the surface \citep{garcia2010,garcia2013}, displaying a hotter region near the upper boundary, where electron scattering is likely the main source of opacity; and a warmer and less ionized region after a relatively sharp temperature decrease at a particular depth in the slab, where the gas recombines rapidly. In this warm and deeper region is where most of the soft X-ray line emission takes place, and photoionization as well as recombination are the leading processes controlling the state of the gas.

The temperature in the slab is determined by the balance of Compton heating and cooling, free–free heating and cooling, photoionization heating, and radiative cooling. Results in Figure~\ref{fig:temperature profiles} show generally good agreements between the two code versions for the densities where they are deemed to be applicable (heating and cooling rates are shown separately in Figure~\ref{fig:heating rates} and \ref{fig:cooling rates}). At high density ($n\gtrsim 10^{20}\,$cm$^{-3}$), free-free heating and cooling become dominant in Model~2, while the thermal equilibrium of Model~1 is still dominated by recombination and emission lines, mostly from hydrogen and helium. This is the direct result of suppressed recombination in Model~2, in which case more free-free cooling/heating will be required to maintain a similar thermal equilibrium solution. In the process of inspecting the heating energy produced by individual elements, we found an issue in the Fe UTA atomic data for Model~1, resulting in incorrect heating rates for iron. Fortunately, iron heating was ignored in the codes used for Model~1, causing $\lesssim 10\%$ systemic uncertainty in the final ionization solution.

\begin{figure*}
\centering
\includegraphics[width=0.8\textwidth]{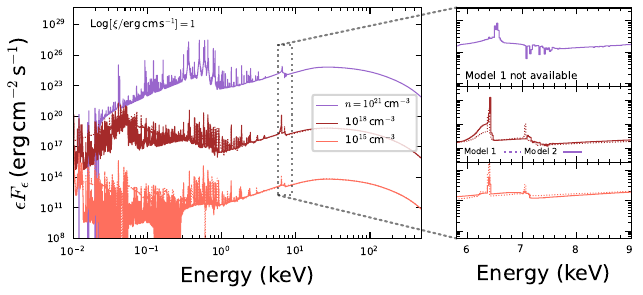}
\includegraphics[width=0.8\textwidth]{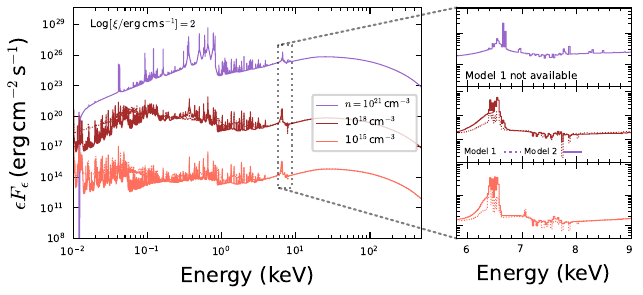}
\includegraphics[width=0.8\textwidth]{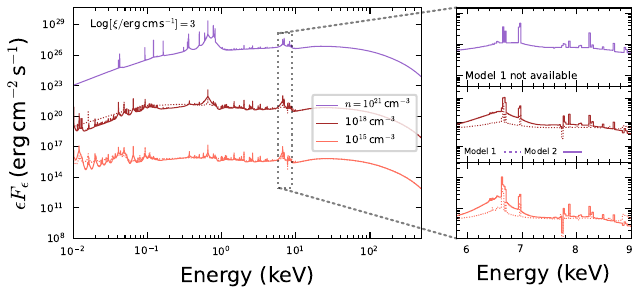}
    \caption{Broad band reflection spectra for different densities and ionizations. Spectra at different densities are normalized by a factor of 1, $10^{2}$, $10^{5}$, $10^{8}$ for clarity. See also Figure~\ref{fig:zoom-in spec} for a clearer view of the narrow atomic features. In the low-density regime, Model~2 is in good agreement with Model~1 except for the narrow line features at low energies. These two versions show larger differences at higher density. The dips present at low energies for models with  $\log(\xi/$erg$\,$cm$\,$s$^{-1})=1$ or 2 are due to the extreme large opacity at those energies, which prevents the radiative transfer solution to fully converge. Model~1 for $n=10^{21}\,$cm$^{-3}$ is not shown because that code cannot properly iterate, due to the incorrect calcium cross-sections.}
    \label{fig:general comparison of spectra}
\end{figure*}

\subsection{Reflected Spectrum}
\label{sec:decomp}

Figure~\ref{fig:general comparison of spectra} shows a comparison of the reflected spectra using the old and new versions of the \xillver\ code (Models~1 and 2), corresponding to the temperature profiles discussed in the previous section. There are significant differences in the spectral features predicted by Model~1 and Model~2. Although the overall spectral shape is similar, there are departures between the two models that are very obvious at soft energies, and the difference becomes more prominent at higher densities. In the iron K region, the relative contribution of resonance emissions from Fe \textsc{xxv} and \textsc{xxiv} are more important for all densities when $\xi=10^3$\,erg\,cm\,s$^{-1}$, which is the typical ionization state for the inner accretion disk. We notice that the suppressed iron edge towards higher density reported in \citet{garcia2016effects} is not observed in the present models. Another important difference to note is that the public \xillver\ tables did not include nickel (i.e., its abundance was set to zero), while here we always take nickel into consideration when comparing \xillver~3.4b and 3.5.

Model~2 predicts more absorption and less recombination emission below $1\,$keV, resulting in lower continuum flux for almost all densities. The flux is not missing but redistributed to the stronger atomic lines around 1\,keV and iron edge. However, there are indeed up to $\sim$10\% discrepancies between input and output flux in high density models, which is likely to stem from a convergence problem of iteration procedure of the radiative transfer solution. The large dips observed
below $\sim 1$\,keV in the spectrum for the lowest ionization and density (top-left panel in Figure~\ref{fig:general comparison of spectra}), are another demonstration of this problem. They are due to the extremely large opacity and  steep down turn of the temperature profile, making the radiative transfer iterations difficult to converge. This can be increasingly problematic as the optical depth become high (i.e. going deep into the disk), and the absorption and emission from low ionization state atoms become important. This issue can be alleviated by increasing spatial sampling rate (e.g. using 1000 layers instead of 200) but a complete solution would require adaptive grid that increases the spatial resolution around the steep temperature cliff and/or switch to more advanced numerical techniques, such as accelerated $\Lambda$-iteration. A careful exploration of these numerical aspecs will be featured in future work.

Based on the comparisons presented here, we can identify two key differences between Model~1 and Model~2: 
\begin{enumerate}
    \item At high density, the continuum flux below $\sim 2\,$keV becomes lower than before, and there is a larger number of emission lines. These effects are linked to the change of recombination calculation in new version. DR suppression and the revised recombination routines in \xstar\ changed the corresponding rate significantly, thus the recombination continuum flux in soft band becomes different. In general, the seemingly stronger continuum absorption in Model~2 is largely from the suppression of radiative recombination from electron continuum (see Section~\ref{sec:decomp}), while a shift in temperature also plays a role.  The inner layers ($\tau_{T}\gtrsim1$) have different temperature profiles between models, with the Model~2 resulting in lower temperatures and showing a transition from the hot to the warmer phases. However, these differences will only be significant when interpreting atomic lines below $\lesssim 0.5\,$keV.
    
    \item When the ionization parameter is high (typical for inner accretion disk, $\xi\sim10^{3} \rm\, erg\,cm\,s^{-1}$), there are significant discrepancies between Model~2 and Model~1 even at the lowest densities considered (Figure~\ref{fig:general comparison of spectra}). This means that the change in the atomic quantities has an impact on the results throughout the whole density range we have considered. Most noticeable is the stronger broad Compton shoulder of iron K complex and much stronger resonance lines. In Model~2, the temperature profile shifts only negligibly, the ion fractions are also quite similar (Figure~\ref{fig:Iron fractions} and \ref{fig:Oxygen fractions}). However, different atomic level solutions are reached in the new calculations, resulting in the buildup of excited level populations \cite[already reported in][]{kallman2001photoionization}. These new solutions feature stronger (primarily resonance) emissions from highly ionized species \citep{kallman2001photoionization}. These lines are Compton and resonantly scattered, producing a long shoulder extending to even $\sim4\,$keV decorated by narrow lines connected to highly ionized iron and oxygen (e.g., Fe~{\sc xxv} at $\sim$6.7\,keV, O~{\sc viii} at $\sim$0.65\,keV). The resulting Compton shoulder is similar to and may change the shape of the relativistic iron K$\alpha$ tail.
    
\end{enumerate}

\begin{figure}
    \plotone{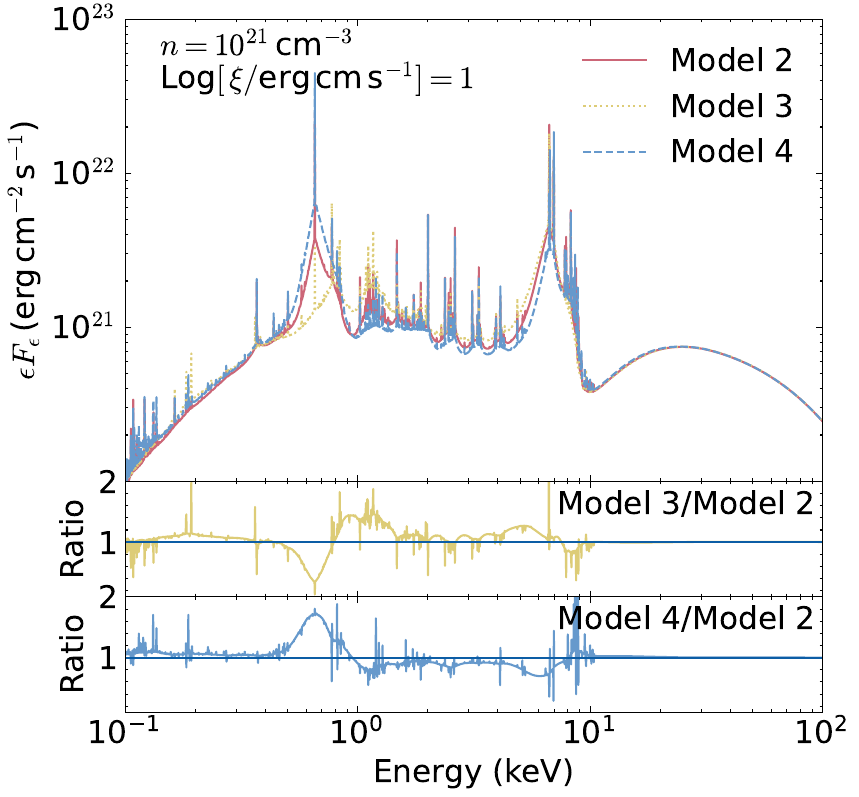}
    \caption{Comparison of various modifications made to the \xstar\ routines when including their updated version into the \xillver\ code (i.e., Model~2). In Model~2, radiative excitation (REX) is turned off (which was on in Model~1), while implementing stimulated recombination. The impact of REX is tested through the comparison between Model~2 and 3. Though REX has relatively large effect on the spectra, we find that in a high-density setting, none of these modifications alone can account for the largest deviations from Model~1 (see Table~\ref{tab:code_versions} for differences between models).}
    \label{fig:morecomp}
\end{figure}

As Model~2 shows increasing deviation from Model~1 when the density becomes large, a natural speculation is that one or multiple of the high density processes dominate Model~2's calculation. Thus, we have explored whether there is a dominant process driving the solutions far from Model~1. In order to do so, we experimented with different ATDBs, implementations of REX and stimulated recombination. \xillver\ running with different settings are considered as different models as summarized in Table~\ref{tab:code_versions}. 
Our comparisons are shown in Figure~\ref{fig:morecomp}. One result consistently observed in all the models computed is that stimulated recombination does not appear to be a dominant process in driving changes in the state of the gas or the shape of the reflected spectrum, and thus we will not discuss the inclusion of this effect any further. The accurate treatment of REX is difficult as the spatial sampling rate must be much shorter than the photon's mean free path. The removal of REX affects primarily the resonance lines. Oxygen resonant transitions are the most affected. These results will be further discussed later in Section~\ref{sec:refl-comp}.

\begin{figure*}
    \centering
    \includegraphics[width=1.0\textwidth]{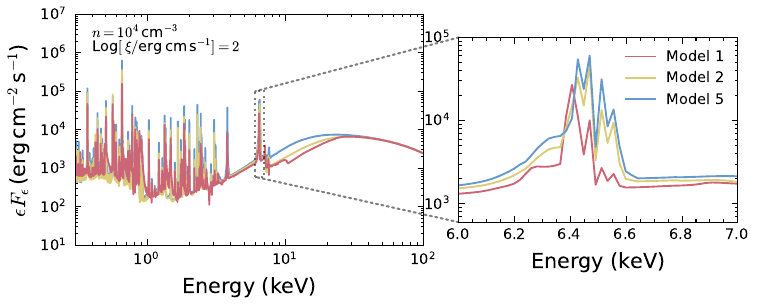}
    \caption{Differences between Model~1, Model~2 and Model~5. In Model~1 and Model~5, the calcium photoionization cross-sections are not corrected. The deviation in 10-20\,keV almost disappears after correcting the cross-section issue. The Fe K line complex shown in the zoom in panel is similar between Model~2 and 5 (notably the line ratio and flux), indicating that the Calcium cross-section issue did not significantly affect the ionization solution.}
    \label{fig:lowncomp}
\end{figure*}

Moderate discrepancy is observed when comparing Model~4 and Model~2. It is worth noting that the high-density effect becomes important earlier than originally anticipated: DR suppression starts to become significant at $n=10^{15}$cm$^{-3}$ \citep{Nikolic2013ApJ...768...82N}. In order to decouple the updates on atomic levels and high density effect, we ran models at a much lower density ($n=10^{4}\,$cm$^{-3}$, Figure~\ref{fig:lowncomp}), a regime in which the DR rates between old and new models should be nearly identical. In this specific comparison, discrepancies in the results revealed the effects of the tabulation error of the photoionization cross-sections of Ca \textsc{xiv} and \textsc{xv}. Though these only affect processes of very low probability, they did result in a factor of a few larger opacity in Model~1 in $\sim10-20\,$keV (see Figure~\ref{fig:lowncomp}) for some ionization parameters. Model~2 handles those highly improbable processes differently than Model~1, creating spurious discrepancies in the Compton hump region. We made the necessary corrections to the atomic database starting from ATDB23. Except for that, the models show good consistency in this low density test.  

\subsection{Fits to Simulated Observations}
\label{sec:modelfit}
One of the most relevant questions in this work is how much the corrections will impact current reflection parameters derived from spectral fitting. The impact can be explored through modeling and simulations of real observations. However, the problem of parameter recoverability, which has been discussed in several works for older models \citep[e.g.][]{bonson2016,choudhury2017,kammoun2018} requires a thorough exploration of the whole parameter space. This is beyond the scope of this work. Therefore, here we limit our discussion to address how different these models can be when it comes to describing typical spectra from accreting BHs.

To illustrate the differences of the models in describing real spectra we produce a series of simulated observations. The simulation is carried out with \xillver~3.5 (specifically Model~2), while the fitting is done with publicly available \xillver, having the same atomic recipe as Model~1. This is done by convolving Model~2 with response matrices and background files derived from authentic \xmm\ and \nustar\ observations on I~Zw~1 \cite[][epoch \textit{b}]{ding2022}. Rather than using the full model, we simplify our test by using the convolutional ray-tracing kernel \textsc{Relconv} \citep{dauser2010,dauser2013,dauser2014} with \textsc{XillverCp}. This is equivalent to treating a small patch on the disk having outgoing photon trajectory characterized by the inclination. 

Unlike our approach in Section~\ref{sec:method}, we use exactly the same input parameters as the public \xillver\ included in \relxill\ version 2.3. Specifically, for the thermal Comptonized coronal spectrum, we set seed black-body temperature $kT_\mathrm{e}=0.01\,$keV; the incident angle is assumed to be 45$^{\circ}$; the inclination of the slab is assumed to be 41$^{\circ}$. We adopt the broken power-law disk emissivity while assuming the inner index, the outer index and the break radius to be 8, 3, and 10 $R_{g}$, respectively, while maintaining solar elemental abundances. The relative strength of the reflection component is controlled by the integrated flux between 20 and 40$\,$keV \cite[reflection strength,][]{dauser2016}, and we subsequently assume integrated flux in reflection component is the same as the thermal Comptonization component in $20-40\,$keV.

The fitting is conducted with the same model suite\footnote{In {\sc xspec}'s terminology: {\tt cflux*relconv*xillverCp+cflux*nthcomp}.}. The disk emissivity profile is very hard to constrain in most cases, basically because the spectrum is not sensitive to it. We consequently froze the outer index and break radius at the injected value, leaving the inner index free to vary. Other parameters are left free to vary, while connecting the photon index and electron temperature between reflection and thermal Comptonization components. 

\begin{figure*}
    \centering
    \plotone{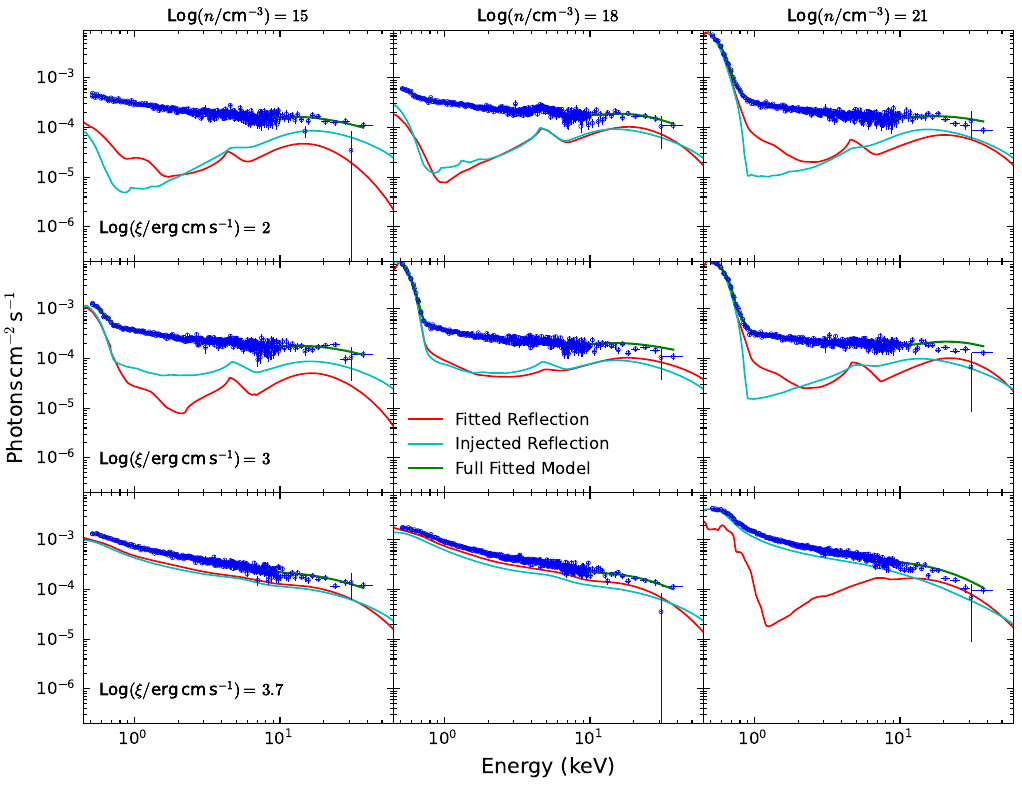}

    \caption{Simulated observations (blue data points), best-fit model (green, power-law + reflection components), injected reflection spectra (cyan, reflection only) and fitted reflection spectra (red, reflection only). Data points are generated by convolving injected spectra (composed of the cyan reflection curve and an unshown power-law, assuming photon index $\Gamma=1.5$.) with instrumental response from \xmm\ and \nustar. Injected spectra refers to the model spectra we assumed when generating the simulated data points. The simulated spectra are fitted with publicly available \xillver\ in \relxill~version 2.3. $\xi$ is the same for each row while densities are the same for each column. Key parameters assumed for each simulation group are shown in the first column and above the first row. This figure illustrates how differently the models describe the same data. In reality, we never know the true reflection curve in astrophysical sources. Here, we can consider the cyan curve as the true value for reflection component.}
    \label{fig:mock-fitting}
\end{figure*}

\begin{figure*}
    \centering
    \plotone{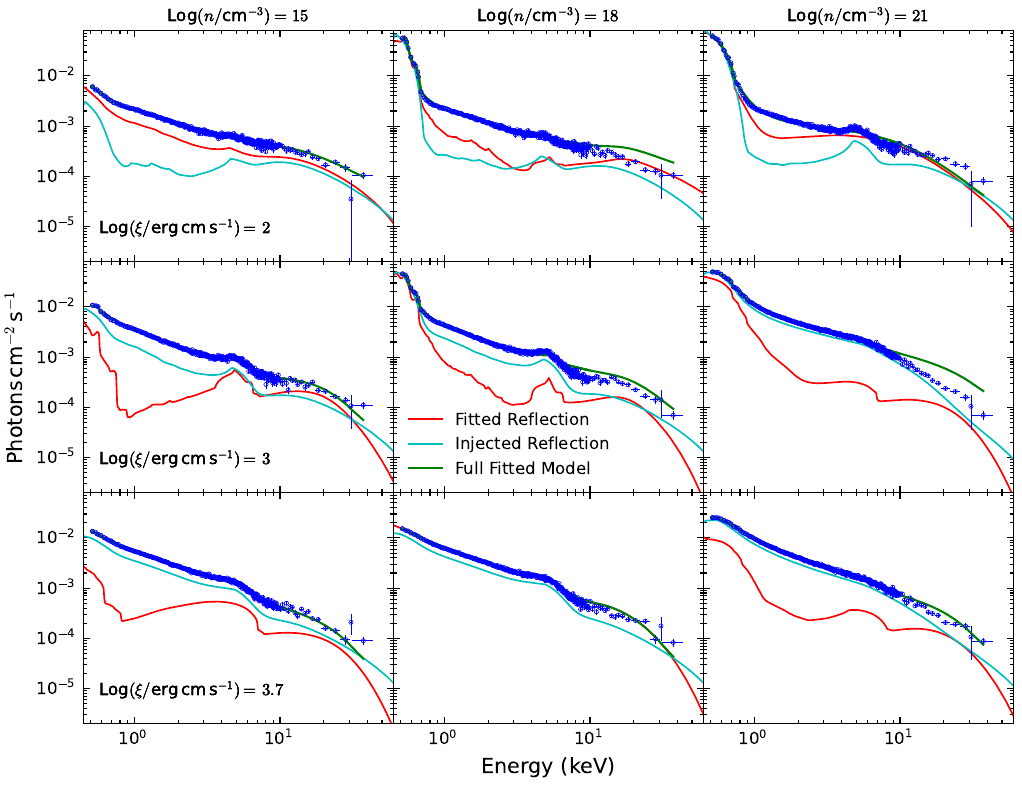}
    \caption{Same as Figure~\ref{fig:mock-fitting} but assuming $\Gamma=2$.}
    \label{fig:mock-fitting_2}
\end{figure*}

Unlike the current public version, \xillver~3.5 uses an updated Compton redistribution scheme \citep{garcia2020accurate}. The impact of different realization of Comptonization is positively correlated with spectra hardness and coronal electron energy ($kT_\mathrm{e}$). In our tests the most extreme case is $\Gamma=1.5$. Even in that case, only the spectrum above $\gtrsim40\,$keV is significantly affected. We estimated the error to be $\lesssim5\%$ at 20\,keV and $\lesssim10\%$ at 40\,keV. As those parts of the spectra make negligible contributions to the fitting statistics, we do not expect them to affect our conclusions.

During the fitting process we found the discrepancies become increasingly large as the ionizing continuum become softer. The results of the simulation are shown in Figure~\ref{fig:mock-fitting}.


These two models are more consistent when the ionizing continuum is hard, in which case less flux is put into the soft energy bins. When fitting the mock spectra with public \xillver, there are more difficulties and larger discrepancies for $\Gamma=2.0$. This is expected because the main differences between them are the atomic recipes which greatly affect the soft band flux. However, even for the $\Gamma=1.5$ and $\log(n/\mathrm{cm}^{-3})=15$ group there are still some differences (Figure~\ref{fig:mock-fitting_2}). These are mainly due to failure of \xillver~3.4b to account simultaneously for the Compton hump and soft excess. The increasing inconsistencies between injected and fitted reflection spectra with increasingly softer continuum is related to the DR suppression effect because $\log(n/\mathrm{cm}^{-3})=15$ is already in the high density regime according to \citet{Nikolic2013ApJ...768...82N}. The high density environment results in the strong suppression of recombination rate, which can be a factor of 10 for some elements (see Figure~\ref{fig:general comparison of spectra} and Section~\ref{sec:decomp}). In summary, our preliminary examination shows that in some cases, including soft-band data may result in systemic errors in the measured parameters, though a good fit with $\chi^{2}/\mathrm{DOF}\simeq1$ and low statistic error can still be achieved for both $\Gamma=1.5$ and 2.0 runs.

We failed to get any acceptable fits for $\Gamma=2.5$, if the density is higher than the typical value ($n=10^{15}\,$cm$^{-3}$). We note that for spectra with $n\gtrsim10^{18}\,$cm$^{-3}$ the formation of the very large iron K edge emission makes \xillver~3.4b completely failed to fit the broad band shape (see also Figure~\ref{fig:zoom-in spec}). However, the greatly boosted oxygen and iron K emission at $\sim0.8$ and $6.4\,$keV in \xillver~3.5 may be able to explain some exotic accreting BH X-ray sources without invoking extreme super-solar iron abundance or multiple reflection components. Such strong emission lines cannot be achieved by classical high density models such as \reflionx HD or \xillver~3.4b as they become almost featureless when moving to the high density regime.

One concern rising from the discussion above is the accuracy of previous high density fitting results. At this point we expect the results
will mainly be shifted in ionization parameter ($\xi$, Equation~\ref{eqn:xi}). As shown in Section~\ref{sec:decomp}, driving process in Model~2 is the change of recombination strength, which is directly linked to the electron density in the denominator of $\xi$.  Given the same $\xi$, the solution found by Model~2 may correspond to a shifted $\xi$ solution in Model~1, which may reduce the difference between Model~1 and 2 at a fixed $\xi$. However, a major difficulty here is that we are unaware of any theoretical method that can quantitatively estimate the shifted effective ionization parameter. One numerical way to show this is to have a full model table, which would permit a cross-correlation matching method. This is beyond the scope of current paper. We plan to make detailed comparisons between models in the near future, with relevant datasets and full tables.

\subsection{Comparing the \xillver\ and \reflionx\ models}
\label{sec:refl-comp}

\begin{figure*}
    \centering
    \includegraphics[width=0.48\textwidth]{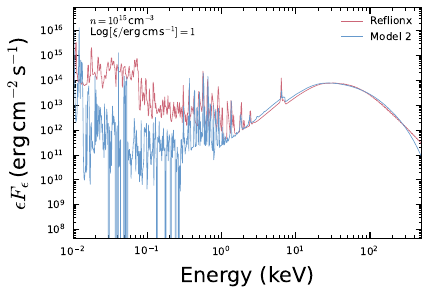}
    \includegraphics[width=0.48\textwidth]{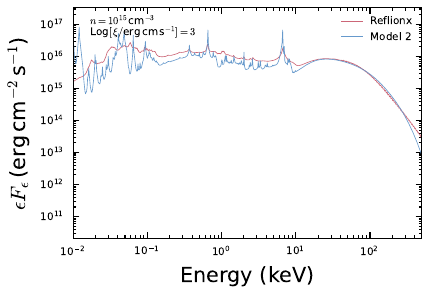}
    \includegraphics[width=0.48\textwidth]{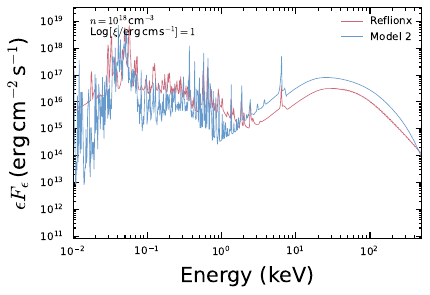}
    \includegraphics[width=0.48\textwidth]{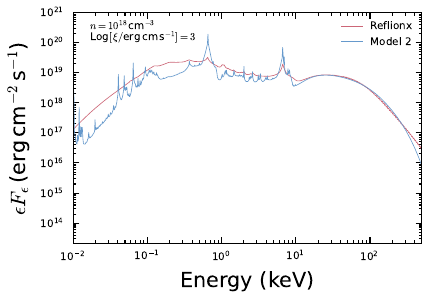}
    \includegraphics[width=0.48\textwidth]{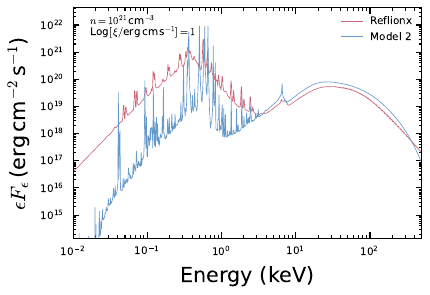}
    \includegraphics[width=0.48\textwidth]{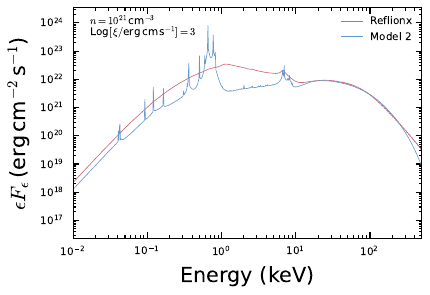}
    \caption{Detailed comparisons between different models. We show \reflionx\ spectra up to its tabulated value $\log(\rm n/erg\,cm\,s^{-1})=21$ for comparison. These models are renormalized such that they have the same integrated flux, corresponding to the quoted ionization parameter and density. When running in high density setting, Model~2 produces stronger iron and oxygen lines in highly ionized environment. Both the \xstar\ routines and the atomic database have impact on the reflection spectrum, with the code producing significant changes in the spectral shape even at high ionization parameters, where atomic emissions were thought to be less important. The lower flux in the soft band is predominantly the result of lower recombination continuum flux rather than stronger absorption. This is due to the shift of ionization equilibrium solution and the energy is redistributed into the hard X-ray emission lines.}
    \label{fig:zoom-in spec}
\end{figure*}

To better illustrate the differences of our model with other codes, we plot the reflection spectra generated by \reflionx~and Model~2 in Figure~\ref{fig:zoom-in spec}. \reflionx\ solves the transfer of the continuum photons using the Fokker–Planck diffusion equation, while implementing the escape probabilities approximation in the calculation of line transfer \citep{ross2005}. \reflionx\ works as a grid of reflection spectra that can be implemented in {\sc xspec} to perform statistical modeling of X-ray spectra (also referred to as a table model). Similar to \xillver, \reflionx\ went through several major updates during last decade. The most recent one included \textsc{nthcomp} as input continuum, while extrapolating the calculation up to density $n=10^{22}\rm cm^{-3}$. The \reflionx\ table used in this paper is the most recent version: {\tt reflionx\_HD\_nthcomp\_v2.fits}\footnote{\reflionx\ tables are publicly available from the IoA ftp site: \url{ftp://ftp.ast.cam.ac.uk/mlparker/reflionx}} \citep{Jiang2020MNRAS,Connors2021ApJ}.

The main differences between the \xillver\ and the \reflionx\ codes are:

\begin{enumerate}
    \item Lager atomic database in \xillver, with a much larger number of lines per ions.
    \item The solution of the radiation transfer is done solving the Fokker--Planck equation in \reflionx, while \xillver\ implements the Feautrier method with lambda iterations.
    \item The energy shifts of the photons due to Compton scattering in \reflionx\ is treated by solving the Kompaneets equation, while the most recent version of \xillver\ uses the full relativistic scattering cross section.
    \item The density dependence of atomic parameters in \reflionx\ follows the formalism of \cite{Summers1970} which includes treatment of DR suppression, while in \xillver\ we follow the \citet{Nikolic2013ApJ...768...82N} fits for DR suppression. \xillver\ also implements the Debye--H\"uckel approximation for the continuum lowering effect.
    \item Other physics incorporated in the \xstar\ routines (and thus in \xillver) that is not included in \reflionx, such as stimulated recombination.
\end{enumerate}

The comparison with \reflionx\ is important because it allows us to validate our results with a completely independent code, which has many similarities to the one presented in this paper. \reflionx\ is only available in relatively low energy resolution. To make this a fair comparison, we resample both models on the same energy grid, with 1000 logarithmically binned grid points spanning from 0.1\,eV to 1000\,keV.

Although we might expect large differences due to the disparities in the ATDB, it turns out that they produce a surprisingly similar continuum shape from $n=10^{15}\,$cm$^{-3}$ all the way to $n=10^{21}\,$cm$^{-3}$. The deviation from \reflionx\ above 100$\,$keV is due to different implementation of Compton redistribution (Section~\ref{sec:method}). 

Looking into the details of those spectra, there are clear differences. The obvious one is the number of emission lines, which is due to the much larger atomic database in \xillver. The second thing to notice is that although the shapes are similar, \xillver\ systematically predicts a lower continuum than \reflionx, at pretty much all energies below 20\,keV. This is because these two codes have converged to different temperature and ionization equilibrium solutions, resulting in a large offset in recombination emissions from all the abundant elements. The most outstanding case is $\log(n/\rm cm^{-3})=21$ and $\log(\xi/\rm erg~cm^{-2}~s^{-1})=1$.

Another noteworthy feature is that \xillver\ seems to almost always produce brighter Fe K lines and deeper K edge, except at the lowest ionization and density. This is likely to be important in the context of BH spin measurements, and it is likely to impact the iron abundances determined with these models, which have been reported to be unexpectedly large in many sources \citep{garcia2018}. The extreme dominance of the oxygen K lines at high density and ionization is quite remarkable, in contrast with the very little emission predicted by \reflionx.

Some of the differences between the \xillver\ and the \reflionx\ spectra are still not fully understood, especially when the density is high. In an effort to do so, we investigated the impact of the various physical processes newly updated between Model~1 and Model~2, as well as the convergence and energy conservation of Model~2.

\begin{figure}
    \centering
    \plotone{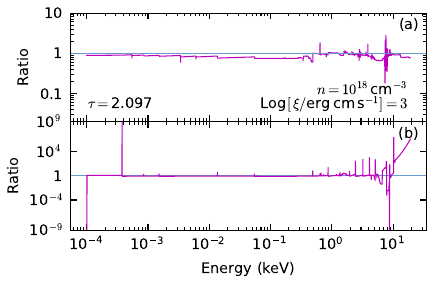}
    \caption{\textbf{(a)} Opacity and \textbf{(b)} emissivity ratio between Model~2 and Model~1, derived immediately after the first global iteration of the code. We show a slice extracted at a specific Thomson depth $\tau\sim2$. \xstar\ has been executed for each of the 200 layers, and 200 $\Lambda-$iterations for the radiative transfer solution have been conducted. A factor of 4$\pi$ has been multiplied into \textbf{(b)} so as to account for the change of internal definition in \xstar.}
    \label{fig:opa_emis}
\end{figure}

The inclusion of REX and/or changes to the ATDB do prominently influence the final results, but they do not contribute to the significant departures in the continuum flux. To understand this better, we looked into the total emissivities and opacities (as a function of energy) calculated by \xstar\ right after the first global iteration in \xillver\ (i.e., the solution of the gas structure at the initial setup) as diagnostics. This is because the radiative transfer iteration couples with the \xstar\ atomic calculation. Results show that most of the differences come from the recombination emissivity, which can vary by a large factor, while opacities are generally within an order of magnitude (Figure~\ref{fig:opa_emis}). The differences in recombination are dominated by DR suppression and the changes in \xstar's code, which becomes more severe with growing density. It is difficult to dissect every change made in \xstar\ over a decade of numerous updates. However, we looked directly into \xstar\ outputs, finding those differences to be minor (factors $\sim$ a few). The coupling between the radiative transfer process and our atomic calculations is likely to be the dominant effect altering the final solutions, getting the slight differences in atomic physics amplified, especially when density is high.

As mentioned in Sec.~\ref{sec:decomp}, large dips are present in the \xillver\ spectrum for the lowest ionization parameter and density (seen at energies below $\sim 0.3$\,keV), which are likely connected with an incomplete or incorrect convergence of the transfer solution.
We have thus tested the energy conservation of Model~2. We find that in all cases we have explored, Model~2 conserves energy better than 5\%, while around 10\% of the total flux escapes the slab through the bottom downward to the center of the disk. We find increasing the spatial sampling from 200 to 1000 improves the convergence of radiative transfer drastically, lowering the energy discrepancy down to below 1\%. Importantly, these un-converged energy grid points have only negligible influence on the X-ray spectrum above 0.1\,keV (less than 1\%). This is because the energy budget is dominated by the emissions in hard X-ray. The oxygen and iron lines in $0.5-7\,$keV can produce up to 80\% of the total flux in the model (in this case, \reflionx\ also has 70\% of the energy in $0.3-10$\,keV). The relatively small energy loss in low energy part generally does not affect the thermal balance of the model. In fact, whether we set 200 or 1000 bins for the reflecting atmosphere, the temperature profiles are almost identical. We also tested separately the output from \xstar\ and find no violation of LTE conditions. 

In summary, \xillver~3.5, specifically Model~2, has passed all sanity checks we have designed, and we are thus confident that it represents the state-of-the-art of reflection modeling.

\subsection{Prospects with \textit{XRISM}}
\label{sec:xrism_test}
High-resolution spectroscopy is the best means to probe the high-density plasma in astrophysical environments, like those discussed here. Successfully lunched on September 6, 2023, X-Ray Imaging and Spectroscopy Mission \citep[XRISM;][]{XRISM2020arXiv}, represents a revolutionary leap forward in X-ray spectroscopy, with a spectral resolution 20-40 times better than the CCD instruments that are used on \textit{Chandra} and \xmm. We have produced simulated XRISM spectra in order to explore the capabilities of these new generation X-ray instruments to detect the small departures in the spectral profiles originated by high-density plasma effects (i.e., differences between Models~1 and 2. See Table~\ref{tab:code_versions}). To simulate an observation, we convolve the spectrum produced by Model~2 with the response files of the {\it Resolve} microcalorimeter in {\it XRISM}, assuming an on-axis point source with an integrated $3-10\,$keV flux of $\sim10^{-11}~\rm erg~cm^{-2}~s^{-1}$. This is the typical flux from cold reflection component in a black hole X-ray binary in the low hard state. We assume a 100\,ks exposure with gate valve closed configuration\footnote{The malfunction of mechanical system on board \textit{XRISM} has resulted in unopened gate valve, which has blocked soft X-ray and incapacitated \textit{Resolve}'s micro-calorimeter under 1.7\,keV. The gate valve also reduces the effective area in all band. The nominal exposure time thus have to increase by, depending on specific scientific objective, as much as $\sim$60\%.}. The simulated spectrum is then fitted by Model~1. 

\begin{figure*}
    \centering
    \plotone{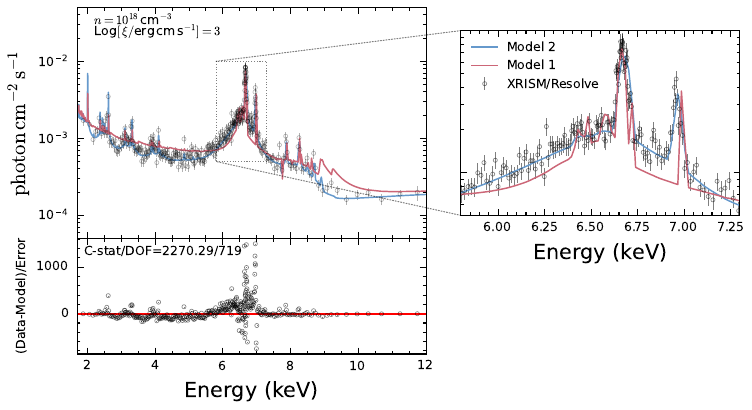}

    \caption{Simulated \textit{XRISM/Resolve} observations (black data points), best-fit with model 1 (red curve) and injected spectrum with model 2 (faint blue). The lower panel shows the fitting statistics and residuals when fitted with model 1. Data points are generated by convolving injected spectrum with \textit{XRISM/Resolve} on-axis point source response files. We assume a 160\,ks exposure with gate valve closed. Key parameters assumed for this simulation are shown on the upper left corner. The photon index $\Gamma$ of the input power-law is 2. The problematic atomic calculations have made model 1 unable to match the iron line complex profile. Note that this fit requires super-solar iron abundance ($A_\mathrm{Fe}\sim5$).}
    \label{fig:xrism}
\end{figure*}

We show the results in Figure~\ref{fig:xrism}. It is evident that a substantial discrepancy exists between Model~1 and the observed data, as indicated by the unsatisfactory best-fit statistic, $C-Stat/DOF = 2270.29/719$. A $A_\mathrm{Fe}\sim5$ super-solar iron abundance is required by the model to properly fit the prominent iron line. As a closing remark, we believe high resolution X-ray spectroscopy will be able to test the various high density models.

\section{Conclusion}
\label{sec:conclusion}
 In this paper, we discussed the necessary corrections required in the atomic data used in the context of accretion disk reflection modeling (Section~\ref{sec:method}). We discussed how the changes in atomic parameters can affect the solution of radiative transfer equations in a plane-parallel slab geometry. The radiative transfer process can amplify the changes in atomic calculations, resulting in different temperature profiles and ionization solutions (Section~\ref{sec:ion-solution}). By comparing different model versions, we find that both the iron K edge and the soft emission lines below 3\,keV are affected by these corrections even in a low density calculation, i.e. $\log(n/\mathrm{cm}^{-3})=15$. We found that DR suppression and updates to the routines that compute the recombination rates are the most important ingredients in the new code (Section~\ref{sec:decomp}). Large discrepancies between \reflionx\ and \xillver\ were observed, and the reason is not yet clear (Section~\ref{sec:refl-comp}). We tested the possible impact of these high-density effects with simulated data. We found the new high density corrections could significantly modify the shape of reflection spectrum in soft X-ray. (Section~\ref{sec:modelfit}). A similar test is conducted in higher spectral resolution observations with simulated {\it XRISM} spectra, which were fitted with the latest public \xillver\ version (Section~\ref{sec:xrism_test}). We predict that strong, narrow resonance lines from highly ionized oxygen and iron atoms, (e.g., O \textsc{vii}, \textsc{viii}; Fe \textsc{xxv}, \textsc{xxvi}) may be evidence of reflection from a high-density disk, although we also note that with some amount of GR blurring such lines may not be recognizable.

 Rather than giving a thorough parameter recoverability test, we provide some general caveats and guidelines for interpreting the parameters in accretion disk reflection modeling:
 
 \begin{enumerate}    
     \item The high-density effects discussed here are particularly evident at soft energies, causing significant departures from the traditional models. Thus, we recommend that particular attention should be given to the sensitivity of the fit parameters in modeling this spectral band.
     
     
     \item The temperature profile we showed in Section~\ref{sec:ion-solution} implies that if the disk itself produces significant flux from the bottom, the ionization solution could be influenced, the result may be especially problematic in the BHXRBs' soft or intermediate states where a strong thermal spectrum dominates the continuum. 

     
     \item Finally, we found that a good fit statistic does not guarantee the correctness of the model parameters. In Section~\ref{sec:modelfit}, we showed that we are still able to get a good fit with current public \xillver\ version. However, the parameters, specifically iron abundance, inclination and density could be biased, sometimes to rather unphysical values. Because the soft X-ray part of those models has such a different parameter space that, the old one converge to totally different ionization solution.
 \end{enumerate}
 
In the near future, a new publicly available full table of models will be made, superseding the current \xillver\ table. This new set of models will extend the range in the density to $10^{21}$ or $10^{22}\,\rm cm^{-3}$ while performing extensive tests on real observations to assess the real impact of the new models, specifically looking at the issue of the Fe abundance.


\section*{Acknowledgements}
We thank James Steiner and Benjamin Coughenour for discussions in the early stage of the paper. YD thanks Luis Ho for inspiring discussions on the future application of the model. This work was supported under NASA Contract No. NNG08FD60C.

All figures in this paper are produced with \textsc{SciencePlots} \citep{SciencePlots}.
\appendix
\section{Heating, cooling rates and Ionization Profiles in the slab}
\begin{figure*}
    \centering
    \includegraphics[width=0.45\textwidth]{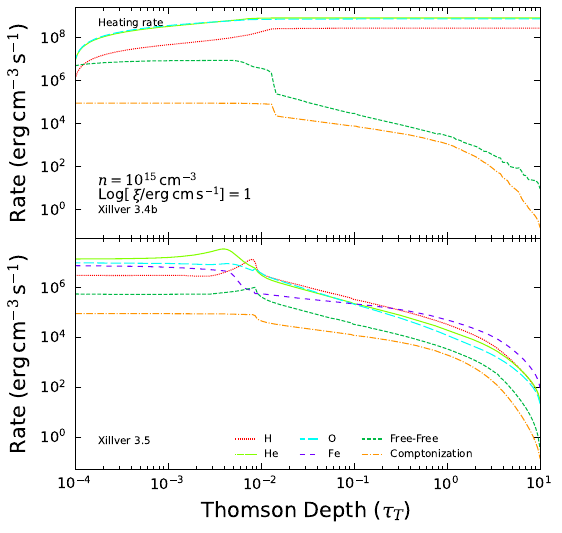}
    \includegraphics[width=0.45\textwidth]{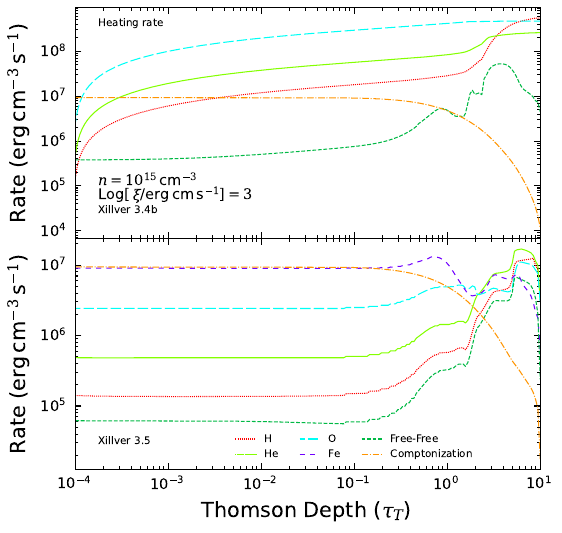}
    \includegraphics[width=0.45\textwidth]{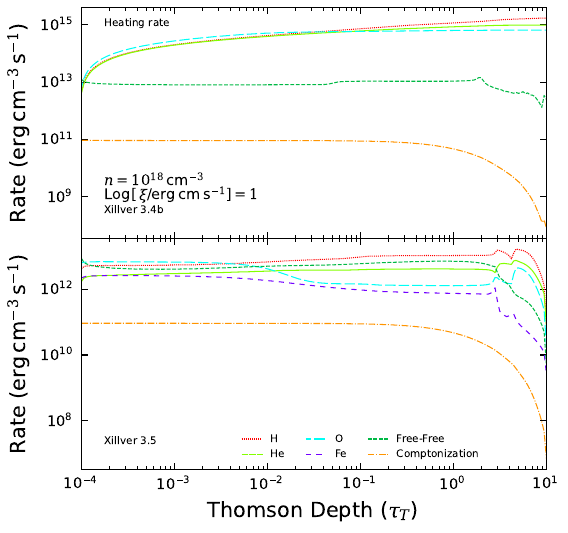}
    \includegraphics[width=0.45\textwidth]{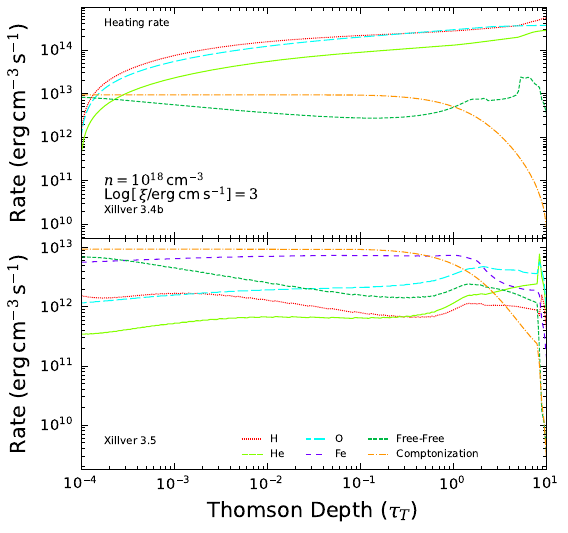}
    \caption{Heating rates corresponding to different densities and ionization parameter $\xi$ as a function of the slab's Thomson optical depth ($\tau_{T}$). Toward higher ionization state and density, free-free heating dominate over other mechanisms quickly. There is an issue in Fe UTA atomic data for \xillver~3.4b, resulting in incorrect heating rates for iron. Iron heating was consequently ignored in that version and may cause $\lesssim 10\%$ systemic uncertainty in the final ionization solution.}
    \label{fig:heating rates}
\end{figure*}

\begin{figure*}
    \centering
    \includegraphics[width=0.45\textwidth]{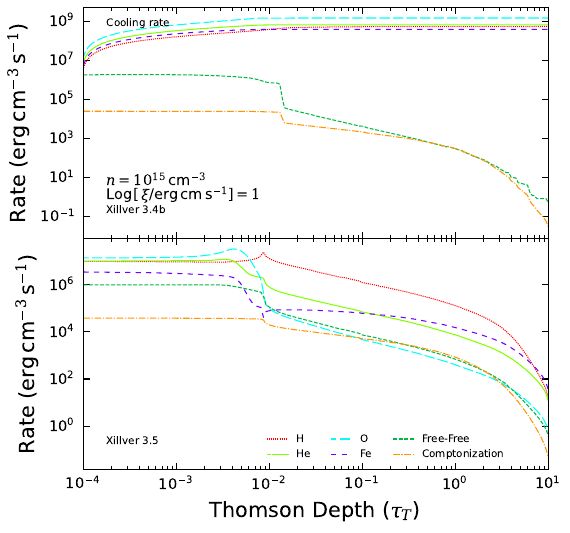}
    \includegraphics[width=0.45\textwidth]{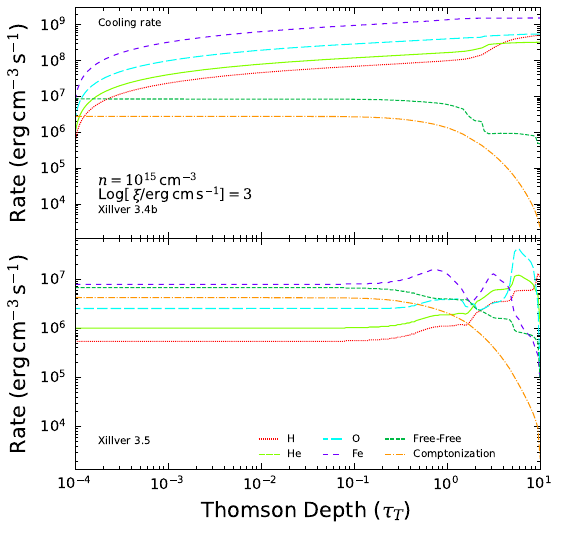}
    \includegraphics[width=0.45\textwidth]{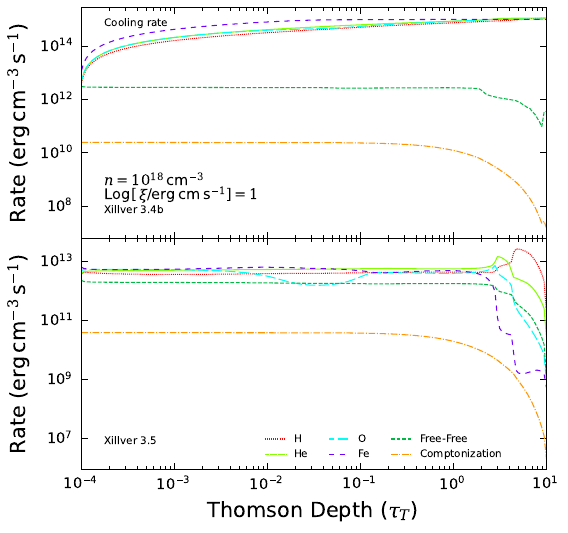}
    \includegraphics[width=0.45\textwidth]{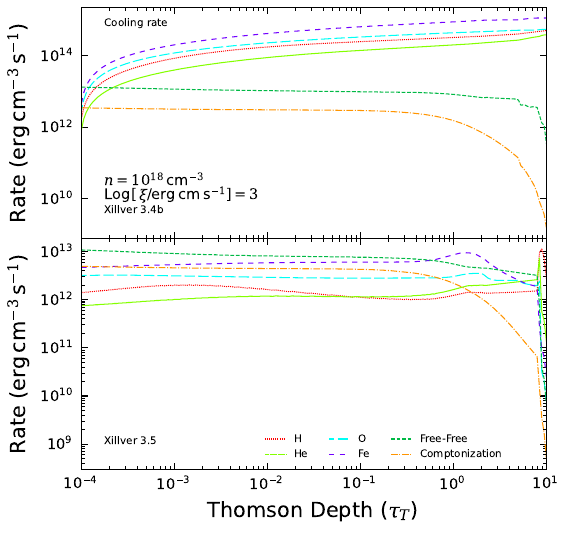}
    \caption{Cooling rates corresponding to different densities and ionization parameters $\xi$ as a function of the slab's Thomson optical depth ($\tau_{T}$). Similar to heating rates, in terms of cooling, free-free and Comptonization also play more important roles in compare with \xillver~3.4b.}
    \label{fig:cooling rates}
\end{figure*}

\begin{figure*}
    \centering
    \includegraphics[width=0.45\textwidth]{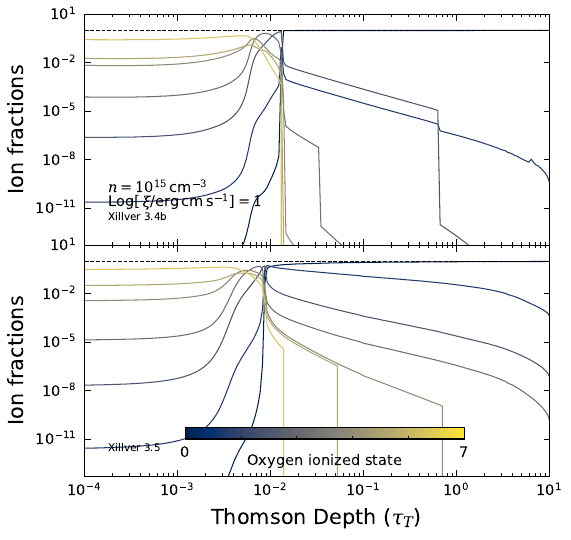}
    \includegraphics[width=0.45\textwidth]{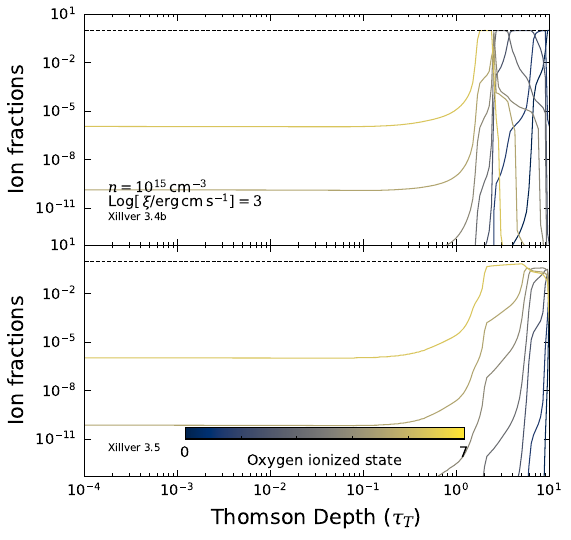}
   \includegraphics[width=0.45\textwidth]{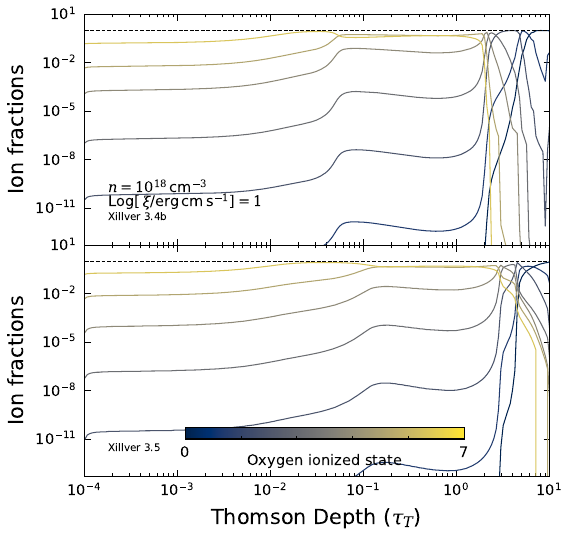}
    \includegraphics[width=0.45\textwidth]{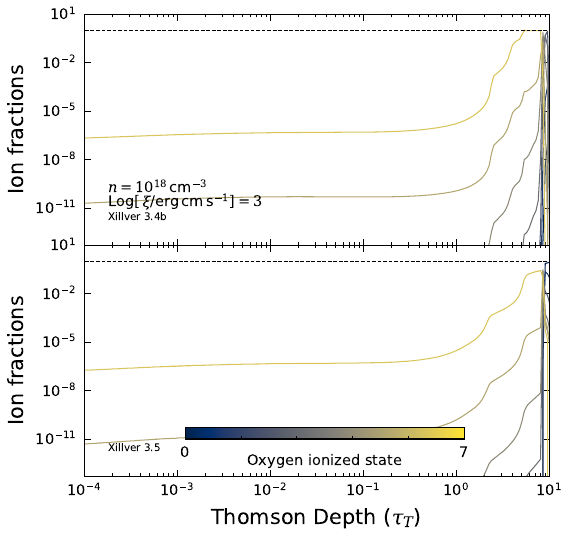}
    \caption{Ion fractions for oxygen corresponding to various densities and ionization parameter $\xi$. In a high density environment, oxygen is less ionized, whose difference is more significant deep inside the atmosphere.}
    \label{fig:Oxygen fractions}
\end{figure*}

\begin{figure*}
    \centering
    \includegraphics[width=0.45\textwidth]{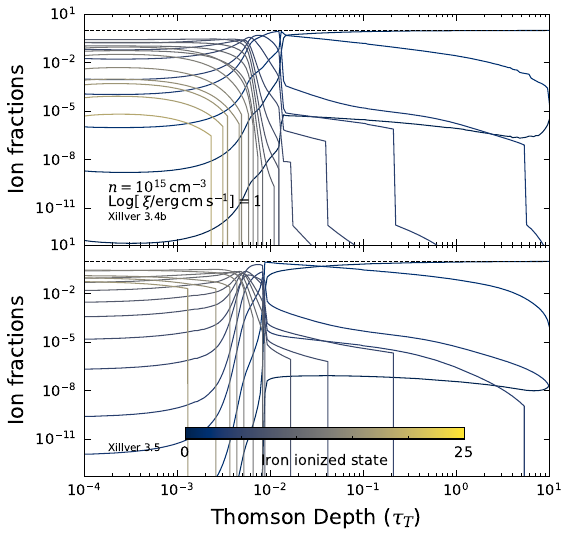}
    \includegraphics[width=0.45\textwidth]{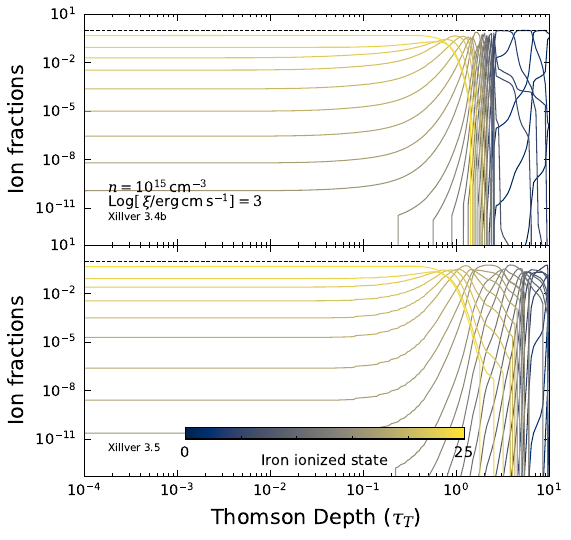}
    \includegraphics[width=0.45\textwidth]{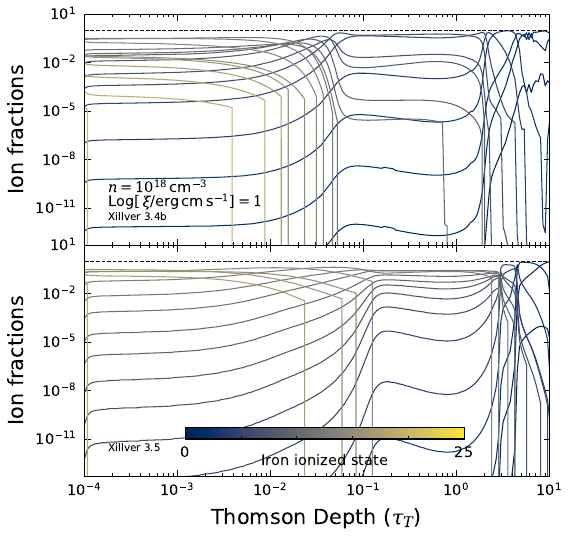}
    \includegraphics[width=0.45\textwidth]{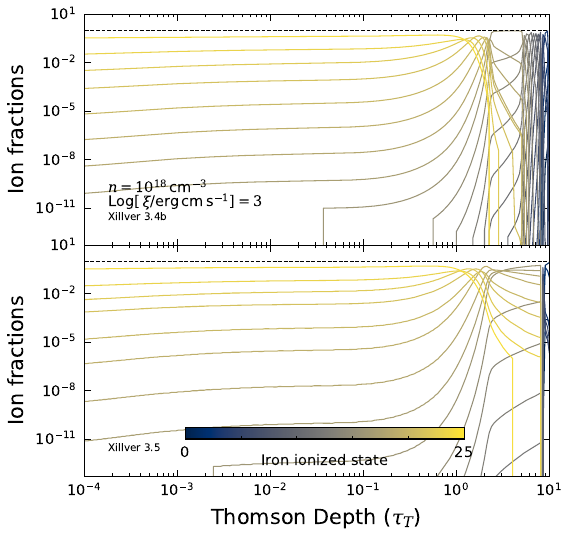}
    \caption{Ion fractions for iron corresponding to various densities and ionization parameter $\xi$, as a function of the slab's Thomson optical depth ($\tau_{T}$). In \xillver~3.5, iron appears to be less ionized in high density environment, though the temperature is almost the same on the surface.}
    \label{fig:Iron fractions}
\end{figure*}

In Figure~\ref{fig:heating rates} and~\ref{fig:cooling rates}, we lists the heating and cooling rates in the solving region, while Figure~\ref{fig:Oxygen fractions} and~\ref{fig:Iron fractions} show ion fractions in different Thomson depth and model settings.
\restartappendixnumbering

\bibliography{my}{}
\bibliographystyle{aasjournal}

\end{document}